\documentclass[aps,pra,twocolumn]{revtex4}
\usepackage{amsmath,amsfonts}
\usepackage[latin1]{inputenc}
\usepackage{graphicx}

\begin{document}
\title{Quantum key distribution using gaussian-modulated 
coherent states}\thanks{Published in 
{\it Nature (London)} {\bf 421}, 238-241 (16 January 2003).}
\date{16 January 2003}

\author{Fr\'ed\'eric \surname{Grosshans},$^a$ Gilles \surname{Van Assche},$^b$ J\'er\^ome \surname{Wenger},$^a$\\ Rosa \surname{Brouri},$^a$ Nicolas J. \surname{Cerf},$^b$ and Philippe \surname{Grangier}$^a$\\~} 
\affiliation{$^a$~Laboratoire Charles Fabry de l'Institut d'Optique, 
CNRS UMR 8501, 91403 Orsay, France\\
$^b$~\'Ecole Polytechnique, CP 165, Universit\'e Libre de Bruxelles,
1050 Bruxelles, Belgium}

\pacs{03.67.Dd, 42.50.-p, 89.70.+c}
\keywords{Quantum cryptography, quantum key distribution, continuous variables, coherent states}

\begin{abstract}
Quantum continuous variables \cite{1} are being explored \cite{2,3,4,5,6,7,8,9,10,11,12,13,14} as an alternative means to implement quantum key distribution, which is usually based on single photon counting \cite{15}. The former approach is potentially advantageous because it should enable higher key distribution rates. Here we propose and experimentally demonstrate a quantum key distribution protocol based on the transmission of gaussian-modulated coherent states (consisting of laser pulses containing a few hundred photons) and shot-noise-limited homodyne detection;  squeezed or entangled beams are not required \cite{13}. Complete secret key extraction
is achieved using a reverse reconciliation \cite{14} technique followed by
privacy amplification. The reverse reconciliation technique is in principle
secure for any value of the line transmission, against gaussian individual attacks based on entanglement and quantum memories. Our table-top experiment 
yields a net key transmission rate of about 1.7~ megabits per second for a loss-free line, and 75~ kilobits per second for a line with losses of 3.1~ dB. We anticipate that the scheme should remain effective for lines with higher losses, particularly because the present limitations are essentially technical, so that significant margin for improvement is available on both the hardware and software.
\end{abstract}
\maketitle

Much interest has arisen recently in using the electromagnetic field amplitudes to obtain possibly more efficient quantum continuous variable (QCV) alternatives \cite{2,3,4,5,6,7,8,9,10,11,12,13,14} to the usual photon-counting quan
tum key distribution (QKD) techniques (see ref. \cite{15} and references therein) --- for instance, by using ``non-classical'' light beams \cite{2,3,4,5,6,7,8,9,10,11}. In fact, it was shown in ref. \cite{13} that squeezed or entangled light is not required to achieve this goal: an equivalent level of security may be obtained by transmitting ``quasi-classical'' coherent states. When the line transmission is larger than 50 \% (line loss $\le$~ 3~dB), the physical limits on QCV cloning \cite{16,17,18} ensure that this protocol is secure against individual attacks. This corroborates the fact that QKD only requires non-orthogonal states, and may well work with macroscopic signals instead of single photons \cite{19}. There are in principle various way
s for the partners Alice and Bob to distribute keys beyond this 3~ dB limit, for instance by using entanglement purification \cite{20} or postselection \cite{12}. Therefore these QCV schemes stimulate many fundamental questions about the physical origin of QKD security. As will be shown below, cryptographic security appears to have a strong relationship with entanglement, even though our protocol does not rely on entangled states.
\par
 
Here we introduce and implement a coherent-state QKD protocol, and we demonstrate that it is, in principle, secure for any value of the line transmission. It relies on the distribution of a gaussian key \cite{7} obtained by continuously modulating the phase and amplitude of coherent light pulses \cite{13} at Alice's side, and subsequently performing homodyne detection at Bob's side. The continuous data are then converted into a common binary key via a specifically designed reconciliation algorithm \cite{8,10}. The security against arbitrarily high losses is achieved by reversing the reconciliation protocol, that is, Alice attempts to guess what was received by Bob rather than Bob guessing what was sent by Alice. Such a reverse reconciliation protocol \cite{14} gives Alice an advantage over a potential eavesdropper Eve, regardless of the line loss. The practical limitations of our scheme are essentially technical, and appear to be due mostly to the limited efficiency of the reconciliation software. 
\par

The protocol runs as follows \cite{13}. First, Alice draws two random numbers $x_A$ and $p_A$ from a gaussian distribution of mean zero and variance $V_A N_0$, where $N_0$ denotes the shot-noise variance. Then, she sends the coherent state $|x_A + i p_A \rangle$ to Bob, who randomly chooses to measure either quadrature $x$ or $p$. Later, using a public authenticated channel, he informs Alice about which quadrature he measured, so she may discard the irrelevant data. After many similar exchanges, Alice and Bob (and possibly the eavesdropper Eve) share a set of correlated gaussian variables, which we call ``key elements''.
\par

Classical data processing is then necessary for Alice and Bob to obtain a fully secret binary key. First, Alice and Bob publicly compare a random sample of their key elements to evaluate the error rate and transmission efficiency of the quantum channel. From the observed correlations, Alice and Bob evaluate the amount of information they share ($I_{AB}=I_{BA}$) and the maximum information Eve may have obtained (by eavesdropping) about their values ($I_{AE}$ and $I_{BE}$). It is known that Alice and Bob can, in principle, distil from their data a common secret key of size $S > {\rm sup}(I_{AB} - I_{AE} , I_{BA} - I_{BE})$ bits per key element \cite{21,22}. This requires classical communication over an authenticated public channel, and may be divided into two steps : reconciliation (that is, correcting the errors while minimizing the information revealed to Eve) and privacy amplification (that is, making the key secret). As we deal here with continuous data, we developed a ``sliced'' reconciliation algorithm \cite{8,10} to extract common bit strings from the correlated key elements. In order to reconcile Bob's measured data with Alice's sent data, the most natural way to proceed is that Bob gets $R$ extra bits of information from Alice in order to correct the transmission errors. The corresponding direct reconciliation (DR) protocols, which have been used so far in QCV QKD \cite{7,13}, allow the generation of a common string of $I_{AB}+R$ bits, of which Eve may know up to $I_{AE}+R$ bits. Here we rather consider reverse reconciliation (RR) protocols \cite{14}, where Bob sends $R$ bits of information to Alice so that she incorporates the transmission errors in her initial data. These RR protocols allow the generation of a common string of $I_{BA}+R$ bits, of which Eve may know $I_{BE}+R$ bits. This turns out to be particularly well suited to QCV QKD, because it is more difficult for Eve to control the errors at Bob's side than to read Alice's modulation. The last step of key extraction, namely privacy amplification, consists of filtering out Eve's information by properly mixing the reconciled bits to spread Eve's uncertainty over the entire final key. This procedure requires an estimate of Eve's information on the reconciled key, so we need a bound on $I_{AE}$ for DR, or $I_{BE}$ for RR. In addition, Alice and Bob must keep track of the information publicly revealed during reconciliation. This knowledge is destroyed at the end of the privacy amplification procedure, reducing the key length by the same amount. The DR bound \cite{13} on $I_{AE}$ implies that the security cannot be warranted if the line transmission $G$ is below 50\%. We will now establish the RR bound on $I_{BE}$, and show that it is not associated with a minimum value of $G$.
\par

In a RR scheme, Eve needs to guess Bob's measurement outcome without adding too much noise on his data. This can be done via an ``entangling cloner'', which creates two quantum-correlated copies of Alice's quantum state, so Eve simply keeps one of them while sending the other to Bob. Let $(x_{\rm in},p_{\rm in})$ be the input field quadratures of the entangling cloner, and $(x_B, p_B)$, $(x_E, p_E)$ the quadratures of Bob's and Eve's output fields. To be safe, we must assume Eve uses the best possible entangling cloner compatible with the parameters of the Alice-Bob channel: Eve's cloner should minimize the conditional variances \cite{23,24} $V(x_B|x_E)$ and $V(p_B|p_E)$, that is, the variances of Eve's estimates of Bob's field quadratures $(x_B,p_B)$. These variances are constrained by Heisenberg-type relations (see Appendix~\ref{Methods}), which limit what can be obtained by Eve:
\begin{eqnarray}
V(x_B|x_A) V(p_B|p_E) \ge N_0^2 \nonumber \\
V(p_B|p_A) V(x_B|x_E) \ge N_0^2
\label{(1)}
\end{eqnarray}
where $V(x_B|x_A)$ and $V(p_B|p_A)$ denote Alice's conditional variances. This means that Alice and Eve cannot jointly know more about Bob's conjugate quadratures than is allowed by the uncertainty principle. Now, Alice's variances can be bounded by using the measured parameters of the quantum channel, which in turn makes it possible to bound Eve's variances.
\par

The channel is described by the linear relations $x_B = G_x^{1/2} (x_{\rm in} + B_x)$ and $p_B = G_p^{1/2} (p_{\rm in}+B_p)$, with $\langle x_{\rm in}^2 \rangle = \langle p_{\rm in}^2 \rangle = V N_0 \ge N_0$, $\langle B_{x,p}^2 \rangle = \chi_{x,p} N_0$, and $\langle x_{\rm in} B_x \rangle = \langle p_{\rm in} B_p \rangle = 0$. Here $\chi_x$, $\chi_p$  represent the channel noises referred to its input, called equivalent input noises \cite{23,24}, while $G_x$, $G_p$ are the channel gains in $x$ and $p$, and $V$ is the variance of Alice's field quadratures in shot-noise units ($V=V_A+1$). The output-output correlations of the entangling cloner, described by $V(x_B|x_E)$ and $V(p_B|p_E)$, depend only on the density matrix $D_{\rm in}$ of the input field $(x_{\rm in},p_{\rm in})$, and not on the way it is produced, namely whether it is a gaussian mixture of coherent states or one of two entangled beams. Inequalities (\ref{(1)}) thus have to be fulfilled for all physically allowed values of $V(x_B|x_A)$ and $V(p_B|p_A)$, given $D_{\rm in}$. Therefore, the values of $V(x_B|x_A)$ and $V(p_B|p_A)$ that should be used in inequalities (\ref{(1)}) to limit Eve's knowledge are the minimum values Alice might achieve by using the maximal entanglement compatible with $V$, namely (see Appendix~\ref{Methods})
\begin{eqnarray}
V(x_B|x_A)_{\rm min} = G_x (\chi_x + V^{-1}) N_0  \nonumber \\
V(p_B|p_A)_{\rm min} = G_p (\chi_p + V^{-1}) N_0
\label{(2)}
\end{eqnarray}
These lower bounds are thus directly connected with entanglement, even though Alice does not use it in practice. They may be compared with the actual values when Alice sends coherent states, that is, $V(x_B|x_A)_{\rm coh} = G_x (\chi_x + 1) N_0$ and $V(p_B|p_A)_{\rm coh} = G_p (\chi_p + 1) N_0$. The lower bounds on Eve's conditional variances are then obtained from equations (\ref{(1)}) and (\ref{(2)}), as:
\begin{eqnarray}
V(p_B|p_E) \ge N_0 / \{ G_x (\chi_x + V^{-1}) \}   \nonumber \\
V(x_B|x_E) \ge N_0 / \{ G_p (\chi_p + V^{-1}) \}
\label{(3)}
\end{eqnarray}
A physical realization of an entangling cloner reaching these bounds is sketched in ref. \cite{14}.
\par

To assess the security of the RR scheme, one assumes that Eve's ability to infer Bob's measurement can reach the limit put by inequalities (\ref{(3)}). For simplicity, we consider the channel gains and noises and the signal variances to be the same for $x$ and $p$ (in practice, deviations should be estimated by statistical tests). The information rates can be derived using Shannon's theory for gaussian additive-noise channels \cite{25}, giving
\begin{subequations}
\begin{eqnarray}
I_{BA} &=& (1/2) \log_2[V_B / (V_{B|A})_{\rm coh}] \nonumber \\
       &=& (1/2) \log_2[( V + \chi ) / ( 1 + \chi ) ]  \label{(4a)} \\
I_{BE} &=& (1/2) \log_2[V_B / (V_{B|E})_{\rm min}] \nonumber \\
       &=& (1/2) \log_2[G^2 ( V + \chi ) ( V^{-1} + \chi ) ] \label{(4b)}
\end{eqnarray}
\end{subequations}
expressed in bits per symbol (or per key element). Here $V_B = \langle x_B^2 \rangle = \langle p_B^2 \rangle = G (V+\chi) N_0$ is Bob's variance, $(V_{B|E})_{\rm min} = V(x_B|x_E)_{\rm min} = V(p_B|p_E)_{\rm min} = N_0 / \{G (\chi + V^{-1})\}$ is Eve's minimum conditional variance, and $(V_{B|A})_{\rm coh} = V(x_B|x_A)_{\rm coh} = V(p_B|p_A)_{\rm coh} = G (\chi + 1) N_0$ is Alice's conditional variance for a coherent-state protocol. The secret bit rate of a RR protocol is thus
\begin{eqnarray}
\Delta I_{RR} &=& I_{BA}-I_{BE} \nonumber \\
              &=& -(1/2) \log_2[G^2 (1 + \chi ) (V^{-1} + \chi )]
\label{(5)}
\end{eqnarray}
and the security is guaranteed if $\Delta I_{RR} > 0$. The equivalent input noise $\chi$ can be split into a ``vacuum noise'' component due to the line losses, given by $\chi_{\rm vac} = (1 - G)/G$, and an ``excess noise'' component defined as $\epsilon = \chi - \chi_{\rm vac}$. In the high-loss limit ($G \ll 1$), the RR protocol remains secure if $\epsilon < (V-1)/(2 V) \approx 1/2$, that is, if the amount of excess noise $\epsilon$ is not too large. In contrast, a DR protocol requires low-loss lines, as the security is warranted only if $\chi < 1$, that is, if $G > 1/(2 - \epsilon)$. Note that DR tolerates an excess noise up to $\epsilon \approx 1$, so it might be preferred to RR for low-loss but noisy channels.
\par

\begin{figure}[t]
\includegraphics{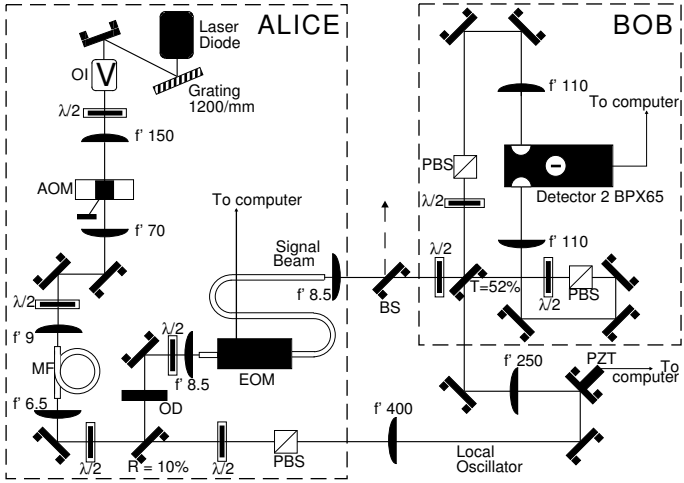}
\caption{Experimental set-up. Laser diode, SDL 5412 (780~nm); OI, optical isolator; $\lambda/2$, half-wave plate; AOM, acousto-optic modulator; MF, polarization maintaining single-mode fibre; OA, optical attenuator; EOM, electro-optic amplitude modulator; PBS, polarizer; BS, beam splitter; PZT, piezoelectric transducer. Focal lengths ($f'$) are given in millimetres. 
$R$ and $T$ are reflection and transmission coefficients.}
\label{fig1}
\end{figure}

\begin{figure}[t]
\includegraphics{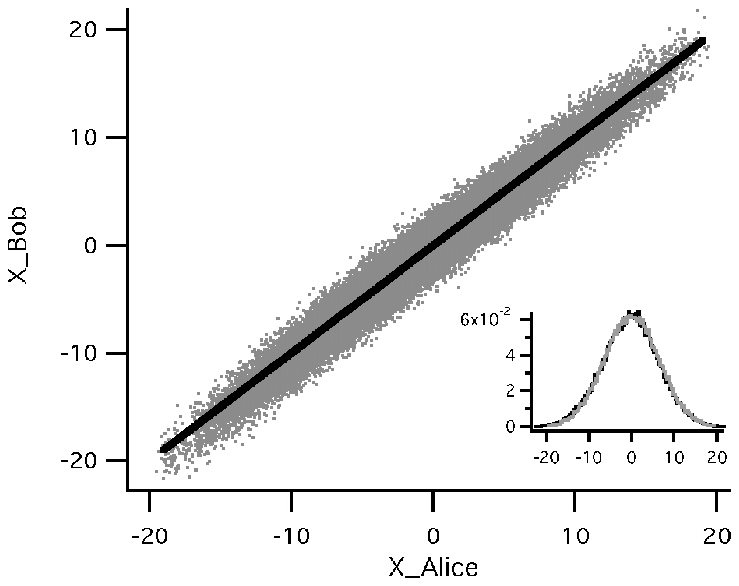}
\caption{Bob's measured quadrature as a function of the amplitude sent by Alice (in Bob's measurement basis) for a burst of 60,000 pulses. The line transmission is 100\% and the modulation variance is V = 41.7. The solid line represents the expected unity slope. Inset, the corresponding histograms of Alice's (grey curve) and Bob's (black curve) data.}
\label{fig2}
\end{figure}

\begin{figure}[t]
\includegraphics{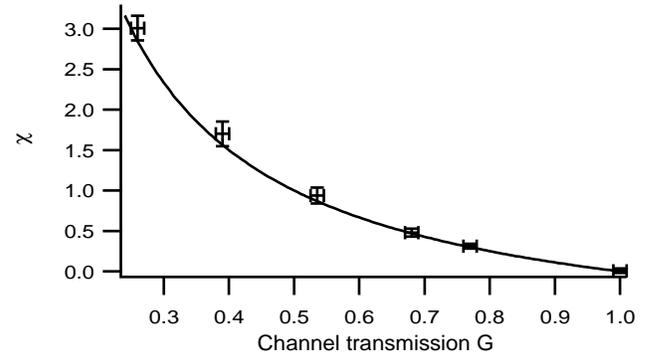}
\caption{Channel equivalent noise $\chi_{\rm line}$ as a function of line transmission $G$. The curve is the theoretical prediction $\chi_{\rm vac} = (1-G)/G$. The errors bars include two contributions with approximately the same size, from statistics (evaluated over blocks of 60,000 pulses) and systematics (calibration errors and drifts).}
\label{fig3}
\end{figure}

\begin{figure}[t]
\includegraphics{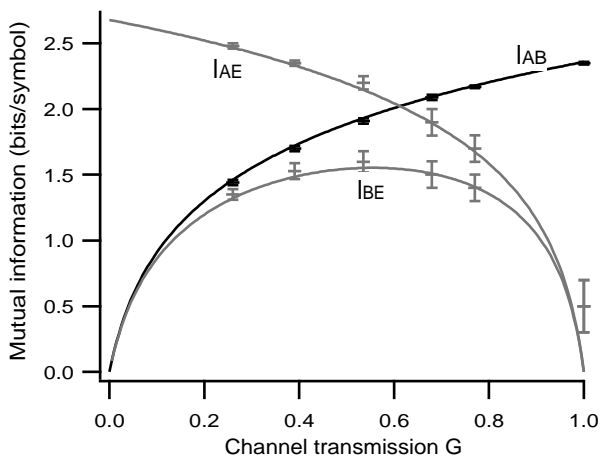}
\caption{Values of $I_{BA}$, $I_{BE}$, and $I_{AE}$ as a function of the line transmission $G$ for $V \approx  40$. Here, $I_{BA}$ is given by equation (\ref{(4a)}), including all transmission and detection noises for evaluating $V_B$ and $(V_{B|A})_{\rm coh}$. The expression for $I_{BE}$ is given by equation (\ref{(4b)}), using the same $V_B$ and  $(V_{B|E})_{\rm min} = N_0 / \{ G (\chi_{\rm line} + V^{-1}) \} + N_{\rm el} + N_{\rm hom}$. This expression realistically assumes that Eve cannot know the noises $N_{\rm el}$ and $N_{\rm hom}$, which are internal to Bob's detection set-up. For comparison with DR, the value of $I_{AE}$ is also plotted (the theoretical value of $I_{AE}$ is obtained from ref. \cite{13}).
}
\label{fig4}
\end{figure}

Our experimental implementation (Fig.~\ref{fig1}) of the quantum key exchange  uses 120-ns coherent pulses at a 800-kHz repetition rate (wavelength of 780~nm, see Appendix~\ref{Methods}). Data bursts of 60,000 pulses have been analysed (Fig.~\ref{fig2}). For each burst, a subset of the values are disclosed to evaluate the transmission $G$ and the total added noise variance. The output noise has four contributions: the shot noise $N_0$, the channel noise $\chi_{\rm line} N_0$, the electronics noise of Bob's detector ($N_{\rm el} = 0.33 N_0$), and the noise due to imperfect homodyne detection efficiency ($N_{\rm hom} = 0.27 N_0$). When introducing line losses using a variable attenuator, the measured $\chi_{\rm line}$ increases as $(1-G)/G$, as shown in Fig.~\ref{fig3} ($\epsilon_{\rm line}= 0$ here). The two detection noises $N_{\rm el}$ and $N_{\rm hom}$ originate from Bob's detection system, so they must be taken into account when estimating $I_{BA}$. In contrast, we may reasonably assume that Eve cannot know or control the corresponding fluctuations, so her attack is inferred on the basis of the line noise $\chi_{\rm line}$ only (see Appendix~\ref{Supplementary_Information} for details). Figure~\ref{fig4} shows explicitly the mutual information between all parties, which makes straightforward the comparison between the DR and RR protocols.
\par

\begin{table*}[t]
\begin{tabular}{llllrrrrrr}
\hline
$V$~~~~~ & $G_{\rm line}$~~ & Losses~~ & $I_{BA}$~~ & $I_{BE}$ & $I_{\rm rec}$ &
~Ideal RR rate & ~Practical RR rate & ~Ideal DR rate & ~Practical DR rate\\
 & & (dB) & (bit) & ~(\% $I_{BA}$) & ~(\% $I_{BA}$) & (kbit~s$^{-1}$) & (kbit~s$^{-1}$) & (kbit~s$^{-1}$) & (kbit~s$^{-1}$)\\
\hline
41.7 &
 1 & 0  &
2.39 &
0 &
88 &
1,920 &
1,690 &
1,910 &
1,660 \\
38.6 &
0.79 & 1.0  &
2.17 &
58 &
85 &
730 &
470 &
540 &
270 \\
32.3 &
0.68 & 1.7  &
1.93 &
67 &
79 &
510 &
185 &
190 &
-- \\
27 &
0.49 & 3.1  &
1.66 &
72  &
78 &
370 &
75 &
0 &
-- \\
43.7 &
0.26 & 5.9  &
1.48 &
93 &
71 &
85 &
-- &
0 &
-- \\
\hline
\end{tabular}
\caption{\textbf{Ideal and practical net secret key rates.} The parameters of the quantum key exchange are measured for several values of the channel transmission $ G$ (the corresponding losses are also given in decibels). The variations of the variance $V$ of Alice's field quadrature are due to different experimental adjustments. The information $I_{BA}$ is given in bits per time slot. Also shown are the maximum information gained by Eve ($I_{BE}$) and the extracted information by reverse reconciliation ($I_{\rm rec}$). The ideal secret key bit rates would be obtained from our measured data with perfect key distillation that yields exactly $I_{BA} - I_{BE}$ bits (RR) or $I_{AB} - I_{AE}$ bits (DR), whereas the practical secret key bit rates are the one achieved with our current key distillation procedure (``--'' means that no secret key is generated). Both bit rates are calculated over bursts of about 60,000 pulses at 800~kHz, not taking into account the duty cycle ($\sim$ 5\%) in the present set-up.}
\label{table1}
\end{table*}

We wrote a computer program that implements the reconciliation algorithm followed by privacy amplification (see Appendices~\ref{Methods} and \ref{Supplementary_Information}). Although Alice and Bob are not spatially separated in the present set-up, the analysed data have the same structure as in a realistic cryptographic exchange. Table~\ref{table1} shows the ideal and practical net key rates of our reverse QKD protocol, as well as the DR values for comparison. The RR scheme is efficient for any value of $G$ provided that the reconciliation protocol achieves the limit given by $I_{BA}$. However, unavoidable deviations of the algorithm from Shannon's limit reduce the actual reconciled information $I_{\rm rec}$ between Alice and Bob, while $I_{BE}$ is of course assumed unaffected. For high modulation ($V \approx 40$) and low losses, the reconciliation efficiency lies around 80\%, which makes it possible to distribute a secret key at a rate of several hundreds of kilobits per second.  However, the achievable reconciliation efficiency drops when the signal-to-noise ratio decreases, but this can be improved by reducing the modulation variance, which increases the ratio $I_{BA}/I_{BE}$. Although the ideal secret key rate is then lower, we could process the data with a reconciliation efficiency of 78\% for $G = 0.49$ (3.1~dB) and $V=27$, resulting in a net key rate of 75~kbits~s$^{-1}$ (see also Appendix~\ref{Methods}). This clearly demonstrates that RR continuous-variable protocols operate efficiently at and beyond the 3~dB loss limit of DR protocols. We emphasize that this result is obtained despite the fact that the evaluated reconciliation cost is higher for RR than for DR: the better result for RR is essentially due to its initial ``quantum advantage''.
\par

In photon-counting QKD, the key rate is limited by the single-photon detectors, in which the avalanche processes are difficult to control reliably at very high counting rates. In contrast, homodyne detection may run at frequencies up to tens of MHz. In addition, a specific advantage of the high dimensionality of the QCV phase space is that the field quadratures can be modulated with a large dynamics, allowing the encoding of several key bits per pulse (see Table~\ref{table1}). Very high secret bit rates are therefore attainable with our coherent-state protocol on low-loss lines. For high-loss lines, our protocol is at present limited by the reconciliation efficiency, but its intrinsic performances remain very high. Since most of the limitations of the present proof-of-principle experiment appear to be of a technical nature, there is still a considerable margin for improvement, both in the hardware (increased detection bandwidth, better homodyne efficiency, lower electronic noise), and in the software (better reconciliation algorithms \cite{26} , see Appendix~\ref{Methods}). In conclusion, the way seems open for implementing the present proposal at telecommunications wavelengths as a practical, high bit-rate, quantum key distribution scheme over long distances.
\par

\paragraph*{Acknowledgements.} The contributions of J. Gao to early stages of the experiment, and of K.  Nguyen to the software development, are acknowledged.  We thank S. Iblisdir for discussions, and Th. Debuisschert for the loan of the 780~nm integrated modulator. This work was supported by the EU programme IST/FET/QIPC (projects ``QUICOV'' and ``EQUIP''), the French programmes ACI Photonique and ASTRE, and by the Belgian programme ARC.
\paragraph*{Correspondence} and requests for materials should be addressed to Philippe Grangier (e-mail {\tt philippe.grangier@iota.u-psud.fr}).

\appendix

\section{Methods}
\label{Methods}

\subsection*{Relevant Heisenberg relations}
In a RR protocol, Alice's estimator for $x_B$ and Eve's estimator for $p_B$ can be denoted respectively as $\alpha x_A$ and $\beta p_E$, where $\alpha$, $\beta$ are real numbers. The corresponding errors are $x_{B|A,\alpha} = x_B - \alpha x_A$, and $p_{B|E,\beta} = p_B - \beta p_E$. Because Alice's, Bob's, and Eve's operators commute, we have $[x_{B|A,\alpha},p_{B|E,\beta}] = [x_B,p_B]$, and thus the Heisenberg relation $\Delta x_{B|A,\alpha}^2 \Delta p_{B|E,\beta}^2 \ge N_0^2$. Defining the conditional variances as $V(x_B|x_A)  = \min_{\alpha}\{\Delta x_{B|A,\alpha}^2\}$ and $V(p_B|p_E) = \min_{\beta}\{\Delta p_{B|E,\beta}^2\}$, we obtain $V(x_B|x_A) V(p_B|p_E) \ge N_0^2$, or, by exchanging $x$ and $p$, $V(p_B|p_A) V(x_B|x_E) \ge N_0^2$.
\par

Alice has the estimators $(x_A,p_A)$ for the field $(x_{\rm in},p_{\rm in}) = (x_A + A_x, p_A + A_p)$ that she sends, with $\langle A_x^2 \rangle = \langle A_p^2 \rangle = s N_0$. Here $s$ measures the amount of squeezing possibly used by Alice in her state preparation \cite{14}, with $s \ge V^{-1}$ for consistency with Heisenberg's relations. By calculating $\langle p_A^2 \rangle = (V - s) N_0$, $\langle p_B^2 \rangle = G_p (V + \chi_p) N_0$, $\langle p_A p_B \rangle = G_p^{1/2} \langle p_A^2 \rangle$, we obtain the conditional variance  $V(p_B|p_A) = \langle p_B^2 \rangle - |\langle p_A p_B \rangle |^2 / \langle p_A^2 \rangle = G_p (s + \chi_p) N_0$. This equation and the constraint $s \ge V^{-1}$ gives $V(p_B|p_A) \ge G_p (V^{-1} + \chi_p) N_0$, and similarly $V(x_B|x_A) \ge G_x (V^{-1} + \chi_x) N_0$. The bound on $V_{B|A}$ is thus obtained by assuming that Alice may use squeezed or entangled beams, while the bound on $V_{B|E}$ can only be achieved if Eve uses an entangling attack. This reflects the fact that squeezing or entanglement play a crucial role in our security demonstration, even though the protocol implies coherent states. Our security proof addresses individual gaussian attacks only, but as the entangling cloner attack saturates the Heisenberg uncertainty relations, we conjecture that it encompasses all incoherent (non-collective) eavesdropping strategies.

\subsection*{Experimental set-up}
A continuous-wave laser diode at 780~ nm wavelength associated with an acousto-optic modulator is used to emit 120-ns (full-width at half-maximum) pulses at a 800 ~kHz rate. The signal pulses contain up to 250 photons, while the local oscillator (LO) power is $1.3 \times 10^8$~ photons per pulse. The amplitude of each pulse is arbitrarily modulated by an integrated electro-optic modulator. However, owing to the unavailability of a fast phase modulator at 780~nm, the phase is not randomly modulated but scanned continuously. No genuine secret key can be distributed, strictly speaking, but random permutations of Bob's data are used to provide realistic data (see Appendix~\ref{Supplementary_Information}). The data are organized in bursts of 60,000 pulses, separated by synchronization periods also used to lock the phase of the LO. The overall homodyne detection efficiency is 0.81, due to the optical transmission (0.92), the mode-matching efficiency (0.96) and the photodiode quantum efficiency (0.92). For the critical data at 3~dB loss, the mode-matching efficiency was improved to 0.99, and thus the overall efficiency was 0.84. We also point out that many blocks of data were exchanged around the 3~dB loss point, with a typical rate above 55 ~kbit~s$^{-1}$.

\subsection*{Secret key distillation}
A common bit string is extracted from the continuous data by sequentially reconciling several strings (``slices'') of binary functions of the gaussian key elements, applying a binary reconciliation protocol successively on each bit \cite{8,10}. Here, we used five  slices, each being corrected either by a trivial one-way protocol (communicating the bits) when the bit error rate (BER) is high, or by the two-way protocol Cascade \cite{27,28} when the BER is low. Note that the disclosed slices are useful for reconciling the remaining slices with less information leaking to Eve, even though they of course do not yield secret bits as such. In addition, Alice and Bob encrypt their classical messages using the one-time pad scheme with a fraction of the previous key bits, or a bootstrap key for the first block. For slices corrected with Cascade, the exchanged parities are encrypted with the same key bits on both sides \cite{29}, making Eve aware of the differences between Alice's and Bob's parities (that is, the error positions) but not of their individual values. Fully communicated slices are also encrypted, thereby revealing no information at all to Eve. Still, Eve may exploit the interactivity of Cascade and gain some information on the final key by combining her knowledge of the error positions with that of the correlations between Alice's and Bob's gaussian values. In the present protocol, this information is numerically calculated for an entangling cloner attack, and then destroyed by privacy amplification. This is achieved by appropriate ``hashing'' \cite{30} functions (see Appendix~\ref{Supplementary_Information}). The resulting net secret key rate is then obtained by subtracting, from the raw key rate, the cost of the one-time pad encryption and the error-position information. Finally, let us emphasize that sliced reconciliation can be made very close to a one-way protocol by increasing the number of key elements from which the bits are jointly extracted (multidimensional reconciliation \cite{8}). This approach was not implemented here, but should deliver an improved secret key rate, approaching the value from the Csisz\'ar-K\"orner formula \cite{21,22}.

\section{Supplementary Information}
\label{Supplementary_Information}

\subsection*{Experimental set-up} 
The source consists of a CW laser diode (SDL 5412) at 780~nm associated with an acousto-optic modulator, used to chop pulses with a duration 120~ns (full width at half-maximum), at a repetition rate 800~kHz. To reduce the excess noise, a grating-extended external cavity is used, and the beam is spatially filtered using a single mode fiber. Light pulses are then split onto a beam-splitter, one beam being the local oscillator (LO), the other Alice's signal beam. The data is organised in bursts of 60,000 pulses, separated by time periods used to lock the phase of the LO, and sequences of pulses to synchronize the parties. In the present experiment, there is a burst every 1.6 seconds, corresponding to a duty cycle of about 5\%, which is obviously under-optimised but should be easy to improve in further experiments.
\par

The coherent state distribution is generated by modulating both the amplitude and phase of the light pulses with the appropriate probability law. In the present experiment, the amplitude of each pulse is arbitrarily modulated at the nominal 800~ kHz rate by an integrated electro-optic LiNbO3 Mach-Zehnder interferometer. In contrast, due to the unavailability of a fast phase modulator at 780~nm, the phase is not randomly modulated but scanned continuously from 0 to $2\pi$ using a piezoelectric transducer (PZT). For such a deterministic phase variation, the security of the protocol is not warranted, and thus no genuine secret key could be distributed strictly speaking. However, the experiment provides realistic data, having exactly the awaited structure provided that random phase permutation on Bob's data are performed.
\par

Due to an imbalance between the paths of the interferometer which modulates the amplitude of the signal beam, the extinction is not strictly zero. In the present experiment that is only aimed at a proof of principle, we subtract the offset field from the data received by Bob. In a real cryptographic transmission, the offset field should be compensated by Alice, either by adding a zeroing field, or by using a better modulator. 
\par

All voltages for the electro-optic modulator or the PZT are generated by an acquisition board (National Instruments PCI6111E) connected to a computer. Although all discussions assume the modulation to be continuous, digitised voltages are used in practice. With our experimental parameters, a resolution of 8 bits is enough to hide the amplitude or phase steps under the shot noise. Since the modulation voltage is produced using a 16 bits converter, and the data is digitised over 12 bits, we may fairly assume the modulation and measurement to be continuous.
\par

The homodyne detection was checked to be shot-noise limited for LO power up to $5\times 10^8$ photons/pulse. In the present experiment, we used $1.3\times 10^8$ photons/pulse for LO power, while each signal pulse contains up to 250 photons. Depending on the run, the overall detection efficiency is either 0.81 or 0.84, due to optical transmission (0.92), mode-matching visibility (0.96 or 0.99) and photodiode quantum efficiency (0.92).
\par
 
The experiment is thus carried out in such a way that all useful parameters can be measured experimentally. Reconciliation and privacy amplification protocols can thus be performed in realistic -- though not fully secret -- conditions. The limitations of the present set-up are essentially due to the lack of appropriate fast amplitude and phase modulators at 780~nm. This should be easily solved by operating at telecom wavelengths (1540-1580~nm) where such equipment is readily available. Let us point out also that it is not convenient to transmit separately the signal and LO, so a better solution would be to use a frequency sideband technique similar to M\'erolla et al. \cite{S1}. Then all light pulses are transmitted together along the same fibre, and a separate radio-frequency is sent from Alice to Bob in order to reconstruct the optical phase information.

\subsection*{Hypothesis about the detector's noise : ``realistic'' vs ``paranoid'' assumptions}

After the quantum exchange, Alice and Bob reveal a subset of their values taken randomly to evaluate the transmission $G$ and the total added noise variance. This variance has four contributions: the shot noise $N_0$, the channel noise $\chi_{\rm line} N_0$, the electronics noise of Bob's detector ($N_{\rm el} = 0.33 N_0$), and the noise due to imperfect homodyne detection efficiency ($N_{\rm hom} = 0.27 N_0$). The two detection noises $N_{\rm el}$ and $N_{\rm hom}$ originate from Bob's detection system, so one may reasonably assume that they do not contribute to Eve's knowledge. This ``realistic'' assumption has been followed in the article. In that case, the noise from Bob's detection system also affects Eve's information so, in equation (\ref{(4b)}), we take $(V_{B|E})_{\rm min} = N_0 / \{G_{\rm line} (\chi_{\rm line} + V^{-1}) \} + N_{\rm el} + N_{\rm hom}$, where $G_{\rm line}$ stands for the line transmission.
\par

In contrast, in a ``paranoid'' approach, one should assume that the noises $N_{\rm el}$ and $N_{\rm hom}$ are also controlled by Eve, that gives her a supplementary advantage. In that case, $(V_{B|E})_{\rm min}$ will be given by $N_0 / \{G ( \chi+ V^{-1})\}$, where $G$ now includes both the line and detection efficiencies and $\chi$ includes both the line and detection noises. In all cases, the value of $I_{BA}$ is given by equation (\ref{(4a)}), where $\chi$ is the total equivalent noise including both transmission and detection.
\par
 
Presently we are able to extract practically a key with up to 3.1~dB losses under the ``realistic'' approach with a reverse reconciliation protocol. Considering now the ``paranoid'' assumption and reverse reconciliation, the ideal secret key rate is 420 kbits~s$^{-1}$ for a lossless line, and 200 kbits~s$^{-1}$ for $G_{\rm line}= 0.79$ (1.0~dB). However, secret bits could be delivered only in the lossless case, at a practical rate of 195 kbits~s$^{-1}$. It is clear that an increase in the reconciliation efficiency would immediately translate into a larger achievable range. Let us point out that we always assume in both the ``realistic'' or the ``paranoid'' approach that Eve has an ideal software, quantum memories, perfectly entangled beams, etc. If any of these hypotheses is relaxed, the practical secure range may be extended over the ``threshold'' presently set by the limited reconciliation efficiency. However, it is not the purpose of the present paper to discuss such ``constrained attacks''.

\subsection*{Implementation of secret key distillation} 

Secret key distillation was performed by a computer program written in standard C++ that implements the steps described in the paper. Although Alice's and Bob's data are both processed on the same computer, it is done in the same way as if the parties were distant and were using a network connection as classical channel. The particular platform used is a regular PC running Linux.
\par

As bursts of data are input to the program, a part of the gaussian key elements are sacrificed and used to estimate the characteristics of the quantum channel. This includes the variances and the correlation coefficient between both sides, which would be exchanged between Alice and Bob over the public authenticated classical channel in a real-life setup. 
\par

Depending on the value of the estimated $I_{AB}$, the two parties agree on appropriate binary functions (slices \cite{8,10}) that will transform their gaussian values into bits. These bits are then reconciled, as described in the paper, with sliced error correction \cite{8,10} and an implementation \cite{28} of Cascade \cite{27} as a sub-routine. In our implementation, 5 binary functions are used per gaussian key elements, out of which 2 or 3 (depending on $I_{AB}$) are fully disclosed, while the remaining 3 or 2 are reconciled using Cascade.
\par
 
Next, the data are moved to the privacy amplification routine. Excluding the bits that are fully disclosed and from which no secret key can be extracted, the reconciled bits are processed by use of a transformation randomly taken in a universal class of hash functions \cite{30,S7}, which in our case is the class of truncated linear functions in a finite field. First, we consider the reconciled bits as coefficients of a binary polynomial in a representation of the Galois field GF($2^{110503}$), hereby called the reconciled polynomial. Then Alice and Bob publicly and randomly choose a random element of the same field and multiply the reconciled polynomial with this chosen element. Finally, they extract from the resulting polynomial the desired number of least significant bits. In our implementation, the representation of the field is GF(2)$[x]/(p)$, where $p = x^{110503}+x^{519}+1$ is an irreducible polynomial over GF(2), see ref. \cite{S8}. The fact that this operation can be implemented efficiently \cite{S9} motivated our choice. The size of the field allows us to process up to 110503 bits at once, or equivalently blocks of about 55200 gaussian key elements when Cascade operates on 2 bits per gaussian key element or of 36800 elements with 3 bits per element. To produce a longer key, the gaussian key elements must thus be grouped into blocks.
\par
 
As explained in the paper, the number of bits that are destroyed by privacy amplification depends on the amount of information that could be inferred by a potential eavesdropper. An easvesdropper Eve has two sources of knowledge. First, she may have attacked the quantum channel and second, she knows the error positions of the reconciled bits from listening to the execution of Cascade. Let $K$ be the final key, $E$ the ancilla Eve uses for quantum eavesdropping, and $\Delta$ the error positions revealed during reconciliation. We thus need to evaluate $I(K;E,\Delta) = I(K;E) + I(K;\Delta|E)$. The first term on the rhs is upper bounded by $I_{BE}$ (in RR) or $I_{AE}$ (in DR), while the second term is evaluated numerically for an entangling cloner attack.
\par
 
This numerical evaluation of $I(K;\Delta|E)$ comes down to integrating $I(K;\Delta|E=e)$ for all possible outcomes $e$ of $E$, weighted by the probability density function $p(e)$ of $E$. In the case of the entangling cloner attack, $E$ refers both to the knowledge of Eve's half of the EPR state she injects and to her eavesdropping of the state being sent to Bob, so $E$ denotes a bivariate gaussian variable whose covariance matrix can be calculated from the channel characteristics (i.e., attenuation and added noise amplitude). For a given outcome $e$ of $E$, Eve can infer $A$ and $B$, Alice's and Bob's key elements as a bivariate gaussian variable. Since $K$ and $\Delta$ are discrete functions of only $A$ and $B$, the probability distribution of $K(A,B)$ and $\Delta(A,B)$ conditionally on $E=e$ can be calculated, hence giving $I(K;\Delta|E=e)$.
\par
 
Finally, a part of the generated key is used to encrypt \cite{29} the execution of the reconciliation for the next block. The remaining bits, namely the net secret key, can be then used for instance to encrypt the classical communications between Alice and Bob using a one-time pad.

\end{document}